# Solar Eclipse Anomalies and Wave Refraction

Alasdair Macleod
University of the Highlands and Islands
Lews Castle College
Stornoway
Isle of Lewis
UK
Alasdair.Macleod@lews.uhi.ac.uk

*Abstract:-* There is some inconclusive evidence that measurement devices sensitive to local gravitation exhibit anomalous behaviour during solar eclipses. We investigate if these findings can be incorporated into the standard general relativistic model of gravitation. The General Theory of Relativity (GTR) describes gravitation as the response of an object to local spacetime curvature. Gravitational waves travelling at the speed of light are then a necessary mechanism to maintain the required consistency between local curvature and distant gravitating mass. Gravitational waves will certainly be subject to refraction by bodies such as the moon and we explore if such an effect can result in an error in the apparent position of the sources and thereby give rise to the characteristic pattern of response associated with the eclipse anomaly. It is found there are phenomenological similarities, but only if gravitational waves are considered not merely to respond to spacetime curvature but are also significantly affected by the presence of mass, perhaps in a manner analogous to electromagnetic waves propagating through matter.

## I Introduction

There is a growing body of evidence supporting the view that changes in gravitational force occur on the surface of the Earth during a solar eclipse of a nature that deviates significantly from that expected merely through the superposition of the gravitational influences of the Sun and the Moon at alignment. The anomalous effect is apparently only detectable using certain types of instrument operating under unspecified but presumably fortuitous circumstances and may depend on unknown factors, including the elevation and azimuth of the solar eclipse and the angular positions of the Earth and the Moon in their mildly eccentric orbits. There is a frustrating problem of repeatability associated with the effect and an evident dependence on dynamic measurement, giving rise to the suspicion of underlying narrow resonance effects. The data currently available is reviewed by Duif [1] who concludes 'there exists strong data that cannot be easily explained away'. There are however reasonable proposals which attempt to explain the data by the action of atmospheric and seismic disturbances that accompany an eclipse[2] and it may well be that the problem can be resolved through this approach. Nevertheless, in this paper we investigate whether there is any possible explanation for the effect in the context of the standard General Relativistic model of gravitation.

Generally speaking, there is a characteristic pattern to the data (Fig. 1) with an initial change in gravitational acceleration before first contact and the opposite effect following full occultation (although attention should also be given to the related gravimeter measurements of Yang and Wang [3] which are discrepant in a quite different way). The peak magnitude corresponds to a change of acceleration in the general range 1-10 x $10^{-8}$ ms$^{-2}$, though the direction is uncertain. One should note that, if real, the anomaly cannot be the result of a tidal residual, as the implied change in the total gravitational force acting on the Earth is then huge and would have a colossal effect on the orbital trajectory (with accompanied changes to the Earth-Moon system that would be readily apparent from laser ranging). The condition that the effect cannot be tidal in origin severely constrains the form of any theory that may be presented to resolve the apparent anomaly.

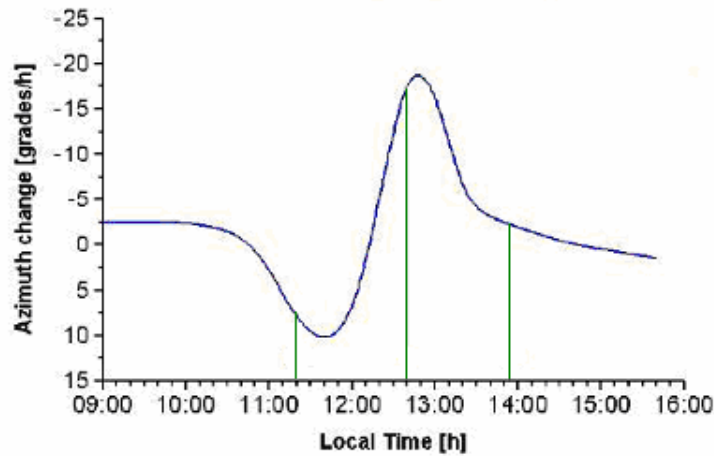

**Figure 1** A recurring pattern emerging in the examples of anomalous data cited by Duif [1] and taken from that paper. This specific example is the smoothed change in the azimuth angle of a paraconical pendulum over the duration of the occultation in the Allias experiment where the effect was first discovered. First contact was at 11:20 h, maximum at 12:40 h, and last contact at 13:55 h. Note the 20-minute deviation from anti-symmetry about the maximum that may support conventional explanations based on atmospheric effects, though most models predict a phase lag rather than the phase lead seen here.

Certainly one proposal to explain the effect (while remaining completely compatible with GTR) is to postulate that local spacetime curvature moves out of step with distant gravitating mass because intervening matter interferes with the mechanism by which local spacetime curvature is updated. This mechanism is believed to be the propagation of gravitational waves at the speed of light, and although these have not been conclusively detected, we can certainly argue for the existence of gravitational waves from the mathematical formulation of GTR and by analogy with the electromagnetic force: The guiding waves of the electromagnetic force are undoubtedly influenced by both spacetime curvature and matter and it is unsurprising to see the apparent position of an object altered by a mirror or lens. We also expect interference effects with optical systems when there is more than one path from source to observer (and the coherence length is not exceeded). Could gravitational waves be similarly affected by matter? From the variability of reported gravitational effects during solar eclipses, and strange oscillatory behaviour uncorrelated with the periodicity of any plausible large-scale physical influence[4], it is almost mandatory that the anomalous gravitational effects reported, if representative of a real effect, should have their origin in an interference phenomenon. It is therefore natural to investigate whether the propagation of gravitational disturbances is more affected by matter than is indicated by the low coupling constant apparent from GTR. An interference process certainly explains the most puzzling aspect of the gravitational anomaly – why the motion of the Earth along its orbital path is unaffected. If the characteristic wavelength of the pattern is much smaller than the size of the Earth (but large enough to be discriminated by laboratory experiments), the interference process will be smeared out and the Earth will simply respond to the averaged effect.

However, one should not give the impression that the *physical* process by with gravitational influences propagates is completely understood, and indeed some confusion even surrounds the physical nature of the electromagnetic force. The mathematical descriptions of gravitation and electromagnetism do not offer any genuine clarification, but do identify significant differences between the two forces; differences that make the development of models merely from analogy quite unsafe. It is therefore worth digressing to see if we can gain some understanding what is meant by electromagnetic and gravitational waves. Immediately excluded from the discussion is any type of hypothetical wave that does not emerge naturally from the formalism of GTR, like for example the guiding waves introduced by some to explain a curious pattern in the spacing of planets empirically described by the Titius-Bode law[5].

**II Electromagnetic and Gravitational Waves**

A body does not act instantaneously on another over distance; instead the influence is subject to a propagation delay proportional to the speed of the mediating quanta – this is considered to be the speed of light for all forces (because of the common propagation medium, spacetime). Whilst the delay could be predicted to cause significant retardation effects, in most situations retardation can be ignored. The reason for this with the electromagnetic force is that observers appear to extrapolate from the retarded position of a source towards the instantaneous position using the latest source velocity information available and respond accordingly, whilst masses subject to the gravitational force extrapolate both the velocity and acceleration of the source[6]. Consequently, charged particles must be accelerated and gravitational systems must be subject to a change in acceleration before the apparent and instantaneous positions differ and retardation discrepancies arise. In all other situations, systems can be analysed in a simple way based on instantaneous rather than time-retarded coordinate positions.

The extrapolation procedure is interesting, but not particularly mysterious. Extrapolation is really a mathematical effect that arises naturally when the equations describing the force are solved under the constraint of appropriate boundary conditions. Of course the extrapolation does not involve any calculation from the retarded to instantaneous positions on the part of the observer; the observer merely responds to the local source of influence in whose dynamics the extrapolation is embedded. Problems only arise when we try to associate a physical mechanism with the process; the mathematical description tells what happens, not how it happens, and it has proved impossible to construct a model that agrees in every respect with observation. For example, it is common in the context of the electromagnetic force to adopt the metaphor of field lines tied to the source that move with the same velocity as the source. When the source is accelerated, the field lines 'catch up' with the new velocity by means of a signal moving outwards at the speed of light as illustrated in Fig. 2.

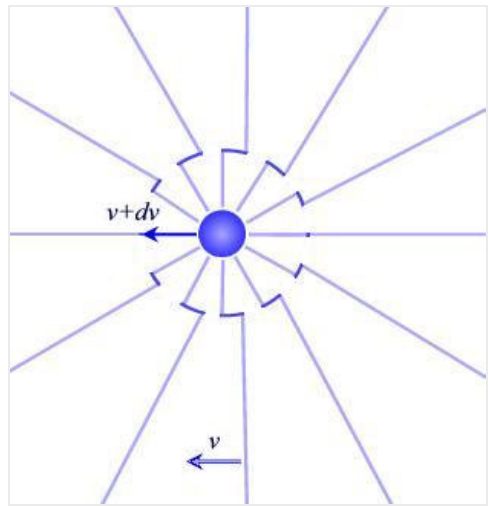

**Figure 2** Field lines representing the instantaneous acceleration of a charge from velocity $v$ to $v + dv$

Note that the diagram might at a glance appear a little misleading – it looks as though the amplitude of the transverse wave increases as the disturbance moves outwards, but there is actually an inverse distance dependence when the drop in field line density with radial distance is taken into account. There is a close similarity between this field line picture and the mathematical wave description emerging from Maxwell's equations, and all observed wave behaviour, including self-interference (as in, for example, the classic twin slit experiment), is fully described.

But in spite of the many successes, the model does not completely match observational data. The propagating wave cannot truly represent the movement of radiant energy in the direct way implied by classical electromagnetism – the electromagnetic effect is quantized, and energy is actually liberated as discrete quanta. In this context the amplitude of the wave at any point should be interpreted as an indication of the probability of a photon being observed (or absorbed) at that position. The question then is whether the wave is real or merely a summary description of how a large ensemble of photons will distribute? This question is still unanswered and possibly unanswerable. In fact when one looks at the process in more detail the situation becomes much more complicated as there is more to electromagnetism than the mere transfer of energy. An electromagnetic disturbance has the secondary purpose of updating the field at all points, and it does so in an apparently non-quantized way and without drawing energy from the wave. This is evident from the examination of a rudimentary example such as a stationary electron subject to the minimum acceleration and surrounded by observers with an ability to measure the local field strength. Only some observers will measure the limited number of synchrotron photons radiated by the gently accelerated charged source, but the field distribution will be updated everywhere and the change will be verified by all observers. Energy transfer certainly represents some sort of direct transaction between an accelerated charged source and observer, but, in contrast, electromagnetic waves stripped of the association with energy transfer are a necessary mechanism to update the field (the electromagnetic 4-potential, $A$). Consider the situation in Fig. 3 (analogous to the solar eclipse situation that is the subject of the paper). No charge is accelerated, yet the force changes. This is easy to understand in terms of the propagation of a field disturbance.

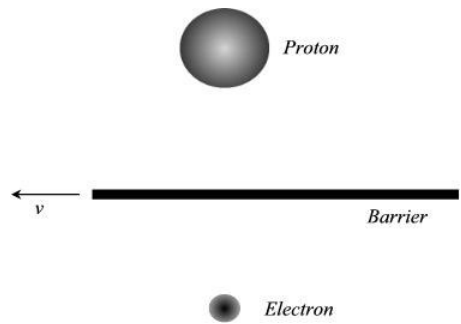

**Figure 3** The barrier is moving at a constant velocity. Will the electron experience an electric and magnetic force from the instantaneous or retarded position of the proton as the barrier moves past?

It is evident that the energy transfer action of the electromagnetic disturbance appears separable from the other function of updating the field, though one should rightly be uncomfortable with incompatible mechanisms for two aspects of what is essentially the same process. The latter role would appear to be completely described by classical electromagnetism and not require quantization. This is really not the case; the electromagnetic force is also completely and effectively described by quantum electrodynamics (QED), which shows the nett interaction to be the sum of many contributing terms. Quantization is then recovered by identifying these terms with the exchange of virtual photons that are individually unobservable and violate the expected energy-momentum relation [7]. The idea of a 'field update' is then understood to be a faulty concept arising merely because we choose to incorrectly impose a field interpretation – in reality, the halo of virtual photons alters with circumstance. Thus, in the situation of Fig. 3, the force alters because the barrier interferes with the flow of virtual photons. Separating this aspect of the operation of the electromagnetic field from energy transfer ultimately fails to allow the quantum character of all interactions to be bypassed by the notion of a field update process. Complete quantization clears many of the conceptual problems running under the surface of the field description given at the beginning of this section, but the relationship between the wave/field and quantum descriptions still remains puzzling, as does the nature of virtual quanta.

Following this rather elaborate detour into electromagnetism, it is now possible to turn back to gravitation and understand that a similar signalling mechanism should exist to sustain the spacetime curvature at points distant from gravitating sources. Although the quantum nature of the gravitational force is not yet understood, the field update mechanism is of necessity the movement of virtual quanta (regardless of any fundamental differences between the electromagnetic and gravitational effects). These virtual quanta are impossible to characterise in the absence of a viable theory of quantum gravity, but they propagate even with no source acceleration and it is possible that information may be gained by investigating wave effects.

**III Refraction by the Moon**

In this section, we wish to consider the possibility that signals (particles) that of necessity propagate through spacetime to ensure the measured gravitational effect far from the Sun is consistent with the extrapolated accelerated position from the viewpoint of the observer are affected by their passage through the Moon. Specifically, we are looking for a refraction effect. The properties of these signals are completely unknown and the only assumption that can sensibly be made is that propagation proceeds at the speed of light appropriate to the local metric.

We will test if an effect analogous to optical refraction occurs because of delays within the medium. We may define a refractive index *n* as the ratio of the speed of the wave phenomenon in the refracting medium (the Moon) to the speed in free space. By undertaking a numerical analysis, the index of refraction that might cause changes in gravity of the character reported will be determined. The characteristic value will then be related to plausible refractive mechanisms in the context of GTR. A ray diagram showing refraction is shown in Fig. 4. The superposition of a direct and refracted signal is postulated to result in an alteration in the strength of the gravitational force at the position of the observer.

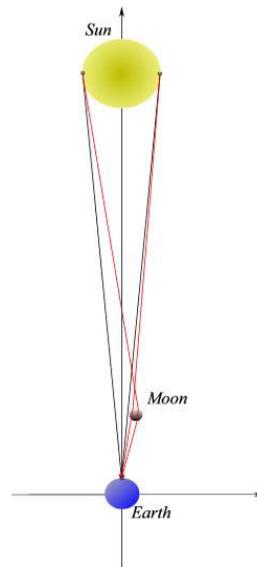

**Figure 4** The diagram (not to scale) shows the bending of what can be loosely described as gravitational waves by the Moon during a solar eclipse. The Moon is moving from right to left relative to the Earth-Sun system. There are two paths, the direct and refracted paths with the expectation of interference at the observation point.

The analysis is generally complex, but because of the large distance between the Moon and the Sun and the huge difference in radii, it is permissible to make some simplifying approximations. Taking the origin of the coordinate system as the position of the observer, the situation is represented in more detail in Fig. 5.

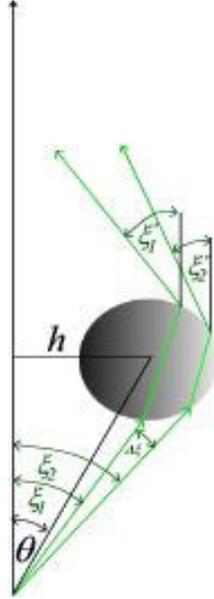

**Figure 5** The angular size of the refracted image is $\Delta\xi$. The leftmost 'ray' originates from the left edge of the sun and the other from the right edge. The direction of the arrows is for convenience of representation and is contrary to the direction of signal movement. It is assumed that the moon is perfectly spherical and of constant refractive index. It is further assumed that the eclipse occurs at the highest point in the sky.

The objective is to determine the angular width of a refracted image of the sun at the origin (as seen by an observer on the Earth conducting gravitational experiments) as it varies with angle $\theta$. The simulation program will be run with $\theta$ ranging from -5 to +5 times the angular width of the Moon in order to cover the entire scale over which experimenters report anomalous data, and a function $\Delta\xi(\theta, n)$ will be numerically derived. It is not possible even to speculate how this signal might be combined with that following a direct path and no attempt will be made to do so. We will merely consider how the horizontal and vertical projections of the refracted image alter with angle. The angular widths of the Sun and Moon are variable, but precision is not important to the analysis and a mean value of 0.009 radians will be adopted.

Let $r_S$ be the mean distance from the surface of the Earth ($E$) to the Sun ($S$), $h$ the perpendicular distance from the Moon to the vector $\overrightarrow{ES}$, $d_{MS}$ the distance from the Moon to the Sun, and $r_S$ the radius of the Sun. Because $r_S \gg h$ over the range of $\theta$ under investigation, the angles $\xi_1'$ and $\xi_2'$ (see Fig. 5) can be treated as constants:

$$\xi_1' \cong \frac{r_S}{d_{MS}} \approx 0.0045 \ rad, \quad \xi_2' \cong -\frac{r_S}{d_{MS}} \approx -0.0045 \ rad. \tag{1}$$

If we let $\Omega_1$ and $\Omega_2$ be the respective angular rotations towards the normal at the surface of incidence, it is clear by symmetry (assuming a homogeneous medium) that

$$\xi_1 - 2\Omega_1 = -\xi_1', \quad \xi_2 - 2\Omega_2 = -\xi_2'. \qquad (2)$$

Let the sine of the angle of incidence (with respect to the normal at the surface) be $\Gamma_i$ where the index takes the values 1 and 2. By elementary trigonometry,

$$\Gamma_i = \frac{d_{EM} \sin(\xi_i - \theta)}{r_M} \qquad (3)$$

where $d_{EM}$ is the distance from the observer to the centre of the moon and $r_M$ is the radius of the moon. The ratio is approximately 221. Because the refractive index is the ratio of the sine of the angles of incidence and refraction, it follows that

$$\Omega_i = \arcsin(\Gamma_i) - \arcsin\left(\frac{\Gamma_i}{n}\right). \qquad (4)$$

Thus to find $\Delta\xi(\theta, n)$ we need to numerically solve the equations

$$\xi_1 - 2\arcsin(221\sin(\xi_1 - \theta)) + 2\arcsin\left(\frac{221\sin(\xi_1 - \theta)}{n}\right) + 0.0045 = 0, \qquad (5a)$$

and

$$\xi_2 - 2\arcsin(221\sin(\xi_2 - \theta)) + 2\arcsin\left(\frac{221\sin(\xi_2 - \theta)}{n}\right) - 0.0045 = 0. \qquad (5b)$$

The range $-0.0045 \leq \theta \leq +0.0045$ defines the region between 1st and 4th contact when the direct route to the sun is progressively obscured

For the purpose of analysis, the refracted 'image' of the Sun is represented as a vector where $\hat{I}$ points in the direction of the centre of the image and $|I|$ is the angular width of the image. Fig. 6 shows a projection of the vector $I$ onto the horizontal axis of an observer viewing an eclipse at the highest point in the sky. Note that first contact occurs at 0.009 radians and last contact at –0.009 radians. There is a clear phenomenological resemblance to the underlying pattern presented in Duif's review[1], and a value for $n$ of 1.005±0.002 seems most representative. Of course, the eclipse is rarely seen in at the highest point in the sky and it might seem that a correction for the true horizontal of the observer is also required. In fact, a rotation of less than one degree would completely destroy the pattern appearing in this simple modelling.

**IV Calculating the Refractive Index**

It does no harm at this stage to emphasise the extreme arbitrariness of the analysis. Yes, it is possible to get curves of the right shape and the desired variability, but it is impossible to make any genuine prediction that can be tested against the operation of a specific instrument for a particular eclipse. This is because the effect of the disruption of signals that convey information about distant gravitating sources on the local measurable gravitational effect is completely unknown. Before even examining this complex issue, it is prudent to consider the plausibility of a refractive index as large as seems to be required. Note that the value of the refractive index is largely unaffected by the details of any theory proposed and is characteristic of any theory that involves refraction rather than alternative concepts such as, for example, gravitational shielding.

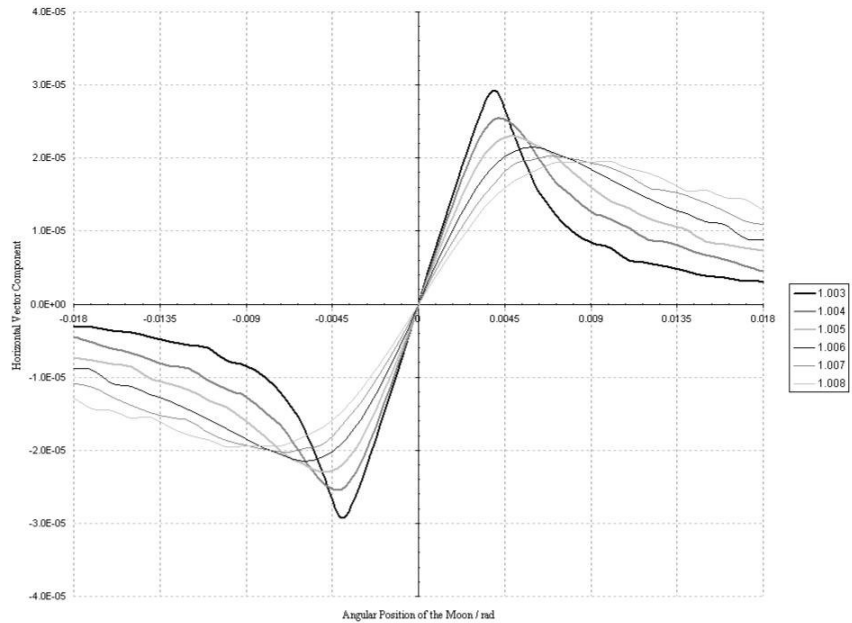

**Figure 6** The horizontal component (i.e. parallel to the surface of the Earth) of the refracted image of the Sun represented as a vector of magnitude $\Delta\xi(\theta, n)$ – *see text*. The shape of the curves for a range of values of refractive index from 1.003 to 1.008 is shown. The horizontal grid is marked in steps equivalent to half the angular width of the Moon.

Of course, like photons, gravitational waves will be slowed and deflected by curved spacetime as they follow a geodesic path. The apparent reduction in propagation speed in a gravitational well with respect to an inertial observer outside the well defines the refractive index. The change in speed can be illustrated by reference to Fig. 7.

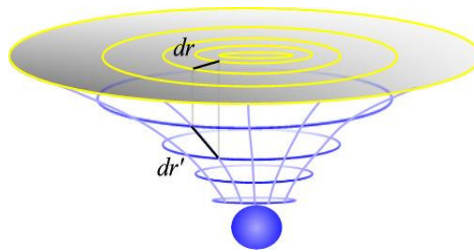

**Figure 7** A graphical representation of the effect of a gravitating source on spacetime curvature. Because a photon follows a geodesic path, the apparent speed of light measured by an external observer will be less than *c*. The measured speed is different for radial and transverse motion.

If $\phi$ is the gravitational potential ($=GM/r$) then the apparent speed of light derived from events that are radially separated is $c\left(1-\dfrac{2\phi}{c^2}\right)$, where the speed is measured locally as *c*. If the experiment instead measures the transverse speed of light (i.e. along an equipotential), this is found to be $c\left(1-\dfrac{2\phi}{c^2}\right)^{\frac{1}{2}}$. If we make the assumption that the potential at the surface of the Moon is constant through the interior (though it actually decreases), then the refractive index is the ratio of the speed of light in free space and within the interior of the Moon. For the purpose of the calculation of the transverse speed equation should be used (because movement is along an equipotential path). Taking the radius and mass of the Moon as 1.74 x $10^6$ m and 7.36 x $10^{22}$ kg respectively, the refractive index is greater than one by an amount 3.13 x $10^{-11}$, a factor of $10^8$ smaller than is required for significant refraction effects.

Of course, we have just described a generalised gravitational lensing effect and it may be that the action of the individual particles cannot be ignored and may have a more significant effect on the update signal as it progresses through the moon (close encountering perhaps $10^{16}$ particles). This is easy to check. Assume the Moon approximates to a body comprising closely packed spheres of radius $R$ (=$10^{-10}$ m) each containing one nucleon at the centre. A photon or gravitational wave will be delayed in passing through the sphere by an amount equal to the integrated delay along the geodesic path. Let the closest distance be $r$. The total delay is simply the Shapiro delay and is equal to [8]

$$\Delta t \cong \frac{4GM_n}{c^3}\ln\left(\frac{2R}{r}\right), \tag{6}$$

where $M_n$ is the mass of a nucleon and t is $2R/c$. The mean fractional delay can be calculated by considering that a signal in proceeding through the Moon is likely to can pass through all the points in the area defined by a circle with radius $R$ with equal probability and weighting (6) accordingly. Thus

$$\overline{\left(\frac{\Delta t}{t}\right)} = \frac{4GM_n}{c^2 R^3}\int_0^R r\ln\left(\frac{2R}{r}\right)dr \tag{7}$$

Integrating between the specified limits,

$$\overline{\left(\frac{\Delta t}{t}\right)} = \frac{4GM_n}{c^2 R^3}\left[\frac{R^2(2\ln(2R/r)+1)}{4}\right]_0^R = \frac{2.38\, GM_n}{c^2 R} \tag{8}$$

For the nucleon mass and typical packing radius, the mean fractional delay is of the order $10^{-44}$, which translates into a refractive index deviation from unity of the same order; completely negligible.

Thus there is no mechanism to explain how a body such as the Moon could impose the level of refraction required, assuming the gravitational update signals are light-like following the curvature of spacetime and are unaffected by matter. The hypothesis that gravitational update signal refraction by the Moon is the cause of the eclipse anomalies cannot be sustained. There is simply no mechanism in GTR for an effect of the correct strength.

**V Discussion**

The basis of this paper is that the update mechanism is a vaguely grey area where anomalous effects *could* possibly lurk. However we have been guided to the view that the gravitational force arises from the exchange of virtual quanta, and the update process for gravitation, though the details are unknown, is severely constrained by a need to correspond to the energy transfer process, gravitational radiation. Gravitational waves interact very weakly with matter, travel at the speed of light and move along geodesics. The update signals must do likewise because the update and transfer processes are expected to operate together when there is change in acceleration. In a sense, radiation and update are two facets of the same process, dynamic and static. If this is correct, there can be no connection between apparent anomalous gravitational effects during eclipses and GTR. One must therefore conclude that the anomalous eclipse effects that have been reported are associated with the terrestrial and atmospheric disturbances accompanying eclipses, particularly so as the paths of artificial satellites (particularly GPS) do not seem to be affected when the Earth frequently occults the Sun and Moon.

However, there exists a troubling paradox: If the gravitational update waves or quanta propagate like photons (the fundamental assumption) then signals cannot escape a black hole, with obvious problems. Moreover, what does happen to virtual quanta that are internally constrained in this way – does pressure increase give rise to energy jets and mass ejection? If black holes do exist, the tidy description of the gravitational interaction (that was constructed by analogy with the electromagnetic force[†]) must be incorrect: The gravitational quantization mechanism then has to be different to what has been suggested. If the alternative process (whatever this may be) should predict a level of interaction with matter to a level comparable with the eclipse effects, one would also anticipate a multitude of anomalous observational and measurement effects associated with rotating bodies such as the Earth and the Sun.

Not least because the role of the gravitational update mechanism of GTR has failed to shed any light on the problem, it is important to really identify the source of the solar eclipse pendulum effects. And because of the Saros cycle – the location pattern of solar eclipses repeats every fifty-four years and one month – the opportunity exists in July 2008 to repeat in every detail the Allias experiment that first revealed the effect.

**References**

[1] C.P. Duif, "A Review of Conventional Explanations of Anomalous observations during Solar Eclipses", gr-qc/0408023 (2004).
[2] T. van Flandern & X.-S. Yang, "Allias Gravity and Pendulum Effects during Solar Eclipses Explained", *Phys. Rev. D* **67** 022002 (2003).
[3] X.-S. Yang & Q.S. Wang, "Gravity Anomaly during the Mohe Total Solar Eclipse and New Constraint on Gravitational Shielding Parameter", *Astrophys. Space Sci.* **282** 245 (2002).
[4] E.J. Saxl & M. Allen, "1970 Solar Eclipse as 'Seen' by a Torsion Pendulum", *Phys. Rev. D* **3** 823 (1971).
[5] P. Lynch, "On the Significance of the Titius-Bode Law for the Distribution of Planets". *Mon. Not. R. Astron. Soc.* **341** 1174 (2003).
[6] S. Carlip, "Aberration and the Speed of Gravity", *Phys. Lett. A* **267** 81(2000).
[7] R. P. Feynman, R.B. Leighton & M. Sands, *The Feynman Lectures on Physics*: *Volume III*, Addison Wesley (1965).
[8] S. Kopeikin, G. Schaefer, A. Polnarev & I. Vlasov, "The Orbital Motion of Sun and a New Test of General Relativity Using Radio Links with the Cassini Spacecraft", gr-qc/0604060 (2006).

---

[†] It is correct to give the static force a quantum interpretation? If we think of the static analogue to the twin slit experiment where the source is a stationary charge, will an observer measuring the electric force along a plane on the other side of the slits record interference effects? In terms of field theory, of course not - there are no waves, the situation is static so there can be no wave effects. However, if the force is really mediated by virtual photons, we would expect these to propagate and self-interfere with a measurable deviation from the prediction of classical field theory. An experiment could technically prove the existence of virtual quanta.